\documentclass[a4paper,12pt]{article}
\usepackage{graphicx}
\usepackage[dvips]{color}
\begin{document}

\parindent=0mm
\parskip 1.5mm

\begin{center}
{\LARGE \bf Analytical relationship for\\ the cranking inertia}

\bigskip

\small \it

Dorin N. Poenaru$^{1,2}$, Radu A. Gherghescu$^1$, and Walter Greiner$^2$
                                                                                
$^1$Horia Hulubei National Instiute of Physics and Nuclear Engineering,
PO Box MG-6, \\077125 Bucharest-Magurele, Romania\\
$^2$Frankfurt Institute for Advanced Studies, J W Goethe University,
Pf. 111932, \\D-60054 Frankfurt am Main, Germany\\
                                                                                
\end{center}
                                                                                
\normalsize \rm

\begin{abstract}
The wave function of a spheroidal harmonic oscillator without spin-orbit
interaction is expressed in terms of associated Laguerre and Hermite
polynomials. The pairing gap and Fermi energy are found by solving the BCS
system of two equations. Analytical relationships for the matrix elements of
inertia are obtained function of the main quantum numbers and potential
derivative. They may be used to test complex computer codes one should
develop in a realistic approach of the fission dynamics. The results given
for the $^{240}$Pu nucleus are compared with a hydrodynamical model. The
importance of taking into account the correction term due to the variation
of the occupation number is stressed.
\end{abstract}

PACS: 24.75.+i, 25.85.Ca, 21.60.-n, 21.10.Pc
                                                                                
\bigskip

\section{Introduction}
By studying fission dynamics \cite{p195b96b} one can estimate the
value of the disintegration constant $\lambda $ of the exponential decay law
expressing the variation in time of the number of decaying nuclei.
The partial decay half-life $T$ is given by  $T=\tau \ln 2 =
0.693147/\lambda$. 
The potential energy surface in a multi-dimensional hyperspace of
deformation parameters $\beta_1, \beta_2, ...., \beta_n$ gives the
generalized forces acting on the nucleus. Information concerning
how the system reacts to these forces is contained in a tensor of
inertial coefficients, or the effective mass parameters $\{ B_{ij}
\}$. Unlike the potential energy $E=E(\beta)$ which depends on the nuclear
shape, the kinetic energy is determined by the contribution of the shape
change expressed by
\begin{equation}
E_k = \frac{1}{2}\sum_{i,j=1}^n B_{ij}(\beta)\frac{d\beta_i}{dt}
\frac{d\beta_j}{dt}
\end{equation}
where $B_{ij}$ is the inertia tensor.
In a phenomenological approach based on incompressible irrotational flow, 
the value of an effective mass  $B^{irr}$ is usually close to the reduced
mass $\mu = (A_1A_2/A)M$ in the exit channel of the binary system. Here $M$
is the nucleon mass. One may use the Werner--Wheeler approximation
\cite{pg197pr95}. 

The microscopic (cranking) model introduced by Inglis
\cite{ing54pr} leads to much larger values of the inertia.
By assuming the adiabatic approximation the shape variations are slower than
the single-particle motion. According to the cranking model, after including
the BCS pairing correlations \cite{bar57pr,bel59kd}, 
the inertia tensor is given by
\cite{bes61kd,bra72rmp}:
\small{\begin{equation}
B_{ij} =2\hbar^2\sum_{\nu \mu} \frac{\langle \nu|\partial H/\partial
\beta_i|\mu \rangle \langle \mu|\partial H/\partial \beta_j|\nu
\rangle}{(E_\nu +E_\mu)^3}(u_\nu v_\mu +u_\mu v_\nu)^2 +P_{ij}  
\label{eq3}
\end{equation}} \normalsize
where $H$ is the single-particle Hamiltonian allowing to determine the
energy levels and the wave functions $|\nu \rangle$, $u_\nu$, $v_\nu$ are
the BCS occupation probabilities, $E_\nu$ is the quasiparticle energy, and
$P_{ij}$ gives the contribution of the occupation number variation when the
deformation is changed (terms including variation of the gap parameter,
$\Delta$, and Fermi energy, $\lambda$,
$\partial \Delta /\partial \beta_i$ and $\partial \lambda /\partial 
\beta_i$):  
\small{\begin{eqnarray*}
\lefteqn{P_{ij} = \frac{\hbar ^2}{4}\sum_\nu \frac{1}{E_\nu ^5} 
\left [ \Delta^2\frac{\partial \lambda}{\partial \beta_i}\frac{\partial
\lambda}{\partial \beta_j} + \right .} \nonumber \\ & &
(\epsilon_\nu - \lambda )^2\frac{\partial \Delta}{\partial \beta_i}
\frac{\partial \Delta}{\partial \beta_j}
 + \Delta(\epsilon_\nu - \lambda )\left(\frac{\partial \lambda}{\partial
\beta_i} \frac{\partial \Delta}{\partial \beta_j} + \frac{\partial
\lambda}{\partial \beta_j} \frac{\partial \Delta}{\partial \beta_i}
\right ) \\ & &
- \Delta ^2 \left ( \frac{\partial \lambda}{\partial \beta_i} \langle \nu | 
\partial H/\partial \beta_j |\nu \rangle +  \frac{\partial \lambda}{\partial
\beta_j} \langle \nu | \partial H/\partial \beta_i |\nu \rangle \right )
\\ & & \left .
- \Delta (\epsilon_\nu - \lambda )\left ( \frac{\partial \Delta}{\partial
\beta_i} \langle \nu |
\partial H/\partial \beta_j |\nu \rangle +  \frac{\partial \Delta}{\partial
\beta_j} \langle \nu | \partial H/\partial \beta_i |\nu \rangle \right )
\right ]
\end{eqnarray*}}  \normalsize

Similar to the shell correction energy, the total inertia is the sum of
contributions given by protons and neutrons $B=B_p+B_n$. The denominator 
in equation (\ref{eq3}) is
minimum for the levels in the neighbourhood of the Fermi energy. A large
value of inertia is the result of a large density of levels at the Fermi
surface. As a result, in a similar way to the shell corrections, one can
observe large fluctuations of $B_{ii}$ when the deformation or the number of
particles are changed. 

In the present work we consider a single-particle model of a spheroidal
harmonic oscillator without spin-orbit interaction for which the cranking
approach allows to obtain analytical relationships of the nuclear inertia.
Despite of the limited interest of this simple single-particle model,
the result of the present work
may be used to test complex computer codes developed in
a realistic treatment of the fission dynamics based on the deformed two
center shell model \cite{wp183302}.
The results illustrated for
$^{240}$Pu nucleus are compared with a hydrodynamical model.

\section{Nuclear shape parametrization}

The shape of a spheroid with semiaxes $a,c$ ($c$ is the semiaxis along the
symmetry) expressed in units of the
spherical radius $R_0=r_0A^{1/3}$ may be determined by a single deformation
coordinate which can be one of the following quantities:
the semiaxes ratio $c/a$; the eccentricity $e=(1-a^2/c^2)$; the deformation
$\delta =1.5(c^2-a^2)/(2c^2+a^2)$, or the quadrupolar
deformation\cite{nil55dk} $\varepsilon = 3(c-a)/(2c+a)$ which will be used
in the following, and according to which the two oscillator frequencies are
expressed as:
\begin{equation}
\omega_\perp (\varepsilon) = \omega_0 \left (1 + 
\frac{\varepsilon}{3} \right )
\end{equation}
\begin{equation}
\omega_z (\varepsilon) = \omega_0 \left (1 - \frac{2\varepsilon}{3} \right )
\end{equation}
and by taking into account the condition of the volume conservation
$\omega_\perp ^2 \omega_z =
({\omega^0_0})^3$ where $\hbar \omega^0_0 = 41 A^{-1/3}$~MeV, one has
\begin{equation}
\omega_0 = \omega^0_0 \left [ 1 - \varepsilon^2\left (\frac{1}{3} +
\frac{2\varepsilon}{27}\right ) \right ] ^{-1/3} 
\end{equation}
A particularly interesting value is $\varepsilon=0.6$ of a superdeformed
spheroid with the ratio $c/a=2$.

\section{Spheroidal harmonic oscillator}

The eigenvalues \cite{p195b96b} in units of $\hbar \omega_0$ are given by
\begin{equation}
E =\hbar \omega _{\perp} (n_{\perp} +1) + \hbar \omega _z (n_z + 1/2) 
 = \hbar \omega _0[N + 3/2 + \varepsilon (n_{\perp} - 2N/3)]
\end{equation}
where the quantum numbers $n_{\perp}$ and $n_z$ are nonnegative integers.
Their summation gives the main quantum number $N=n_{\perp}+n_z$. In units of 
$\hbar \omega _0$ one has a linear variation of the energy levels in
function of deformation $\varepsilon$. By including the variation of 
$\omega _0$ with $\varepsilon$, one obtains the analytical relationship 
\begin{equation}
\epsilon _i = E_i /\hbar \omega^0_0=[N+3/2+\varepsilon (n_{\perp} -
2N/3)][1-\varepsilon^2(1/3+2\varepsilon /27)]^{-1/3}
\end{equation}
in units of  $\hbar \omega^0_0$. Due to the Pauli principle, each energy
level $\epsilon _i$, with quantum numbers $n_{\perp}$ and $N$, can
accomodate $g=2n_{\perp} +2$ nucleons. One has a number of $(N+1)(N+2)$
nucleons in a completely occupied shell characterized by the main quantum
number $N$, and the total number of states for the lowest $N+1$ shells is 
$\sum_{N=0}^{N}(N+1)(N+2)= (N+1)(N+2)(N+3)/3$. For each value of $N$ there
are $N+1$ levels with $n_{\perp} = 0, 1, 2, ... N$. When the deformation 
$\varepsilon > 0$ increases, a level with $n_{\perp} =0$ decreases in energy
and the one with $n_{\perp} =N$ increases. For some particular values of the
deformation parameter there is a crossing of several levels in the same
point leading to a degeneracy followed by an empty gap. If no spin-orbit
coupling is considered for the vanishing deformation parameter, $\varepsilon
= 0$, one has the following sequence of magic numbers: 2, 8, 20, 40, 70,
112, 168, 240, ... If now $\varepsilon = 0.6$ they become 2, 4, 10, 16, 28,
40, 60, 80, 110, 140, 182, ... The known experimental values can be obtained
only with a spin-orbit coupling included.

The spin $\Sigma$ contributes with positive or negative values (up or down)
for every state with quantum numbers $n_z$, $n_r=0, 1, 2, ... n_\perp$ and 
$m=n_\perp -2n_r$, hence in a system of cylindrical coordinates
$(\rho,\varphi,z)$ the wave function \cite{dam69np,vau73prc} can be written
as
\begin{equation}
\Psi=|n_r m n_z \Sigma \rangle = \psi_{n_r}^m (\rho)\Phi_m (\varphi)
\psi_{n_z}(z) \chi(\Sigma)
\end{equation}
Few examples of the quantum numbers $n_r m n_z$ belonging to the lowest
levels with $N=0, 1, 2$ are given in the table~\ref{tab1}. 

\begin{table}[htb] 
\caption{Quantum numbers of the lowest states of a spheroidal harmonic
oscillator}
\begin{center}
\begin{tabular}{|c|c|c|c|c|c|}   \hline
{} &{} &{} &{} &{} &{} \\
{$N$} & {$n_\perp$} & {$n_r$} & {$n_z=N-n_\perp$} & {$m=n_\perp - 2n_r$} & 
{$\epsilon$ for $\varepsilon=0$} \\
{} &{} &{} &{} &{} &{} \\  \hline
 0 & 0 & 0 & 0 & 0 & 1.5 \\ \hline
 1 & 0 & 0 & 1 & 0 & 2.5 \\
   & 1 & 0 & 0 & 1 &   \\ 
   & 1 & 1 & 0 & -1 &  \\  \hline
 2& 0 & 0 & 2 & 0 & 3.5 \\
  & 1 & 0 & 1 & 1 &    \\
  & 1 & 1 & 1 & -1 &    \\
  & 2 & 0 & 0 & 2 &    \\
  & 2 & 1 & 0 & 0 &    \\
  & 2 & 2 & 0 & -2 &    \\    \hline
\end{tabular}\label{tab1}
\end{center}
\end{table}

The eigenfunctions are given by
\begin{equation}
\psi_{n_r}^m (\rho) =
\frac{\sqrt{2}}{\alpha_\perp}N_{n_r}^m\eta^{\frac{|m|}{2}}e^{
-\frac{\eta}{2}}L_{n_r}^{|m|}(\eta) = \frac{\sqrt{2}}{\alpha_\perp}
\psi_{n_r}^m(\eta)
\end{equation}
\begin{equation}
\Phi_m (\varphi) = \frac{1}{\sqrt{2\pi}}e^{im\varphi}          
\end{equation}
\begin{equation}
\psi_{n_z}(z) = \frac{1}{\sqrt{\alpha_z}}N_{n_z} e^{-\frac{\xi^2}{2}} 
H_{n_z}(\xi) = \frac{1}{\sqrt{\alpha_z}}\psi_{n_z}(\xi)
\end{equation}
where $L_{n_r}^{|m|}$ are the associated (or generalized) Laguerre
polynomials and $H_{n_z}$ are the Hermite polynomials. The new dimension-less 
variables $\eta$ and $\xi$ are defined by
\begin{equation}
\eta=\frac{\rho^2}{\alpha_\perp^2}
\end{equation}
\begin{equation}
\xi=\frac{z}{\alpha_z}
\end{equation}
where the quantities
\begin{equation}
\alpha_\perp = \sqrt{\frac{\hbar}{M\omega_\perp}} \approx A^{1/6} \sqrt{
\frac{\omega_0^0}{\omega_\perp}}  \; \; \; \;  \; \; \; \; 
\alpha_z = \sqrt{\frac{\hbar}{M\omega_z}} \approx A^{1/6} \sqrt{
\frac{\omega_0^0}{\omega_z}}
\end{equation}
which depend on the nucleon mass, $M$, posses a dimension of a length. Their
numerical values, in fm, can be estimated by knowing that $\hbar^2/M \approx 
41.5$~MeV$\cdot$fm$^2$ and $\hbar \omega^0_0 = 41 A^{-1/3}$~MeV. 
The normalization constants 
\begin{equation}
(N_{n_r}^m)^2=\frac{n_r!}{(n_r+|m|)!}
\end{equation}
\begin{equation}
(N_{n_z})^2=\frac{1}{\sqrt{\pi}2^{n_z}n_z!}
\end{equation}
are obtained from the orthonormalization conditions
\begin{equation}
\int_0^\infty\psi_{n_r'}^m(\rho)\psi_{n_r}^m(\rho)\rho d\rho = \delta_{n_r'
n_r}
\end{equation}
\begin{equation}
\int_{-\infty}^\infty\psi_{n_z'}(z)\psi_{n_z}(z) dz = \delta_{n_z' n_z}
\end{equation}
\begin{equation}
\int_0^{2\pi}\Phi_{m'}^* (\varphi) \Phi_{m}(\varphi) d\varphi = 
\delta_{m' m}
\end{equation}
One should take into account that the factorial $0!=1!=1$. We shall
substitute the wave functions $\psi_{n_r}^m(\eta)$ and $\psi_{n_z} (\xi)$ in
the equation~(\ref{eq3}) of the nuclear inertia.

\section{Nuclear inertia}

By ignoring the spin-orbit coupling the Hamiltonian of the harmonic
spheroidal oscillator contains the kinetic energy and the potential energy
term, $V$: 
\begin{equation}
V(\eta,\xi;\varepsilon) = \frac{1}{2}M(\omega _{\perp}^2\rho^2 + 
\omega _z^2z^2) =  \frac{1}{2}\hbar \omega _{\perp}\eta +
\frac{1}{2}\hbar \omega _z\xi^2 = 
\frac{\hbar \omega _0^0\left [(3+\varepsilon)\eta + (3-2\varepsilon)\xi^2 
\right ]}{2[27-\varepsilon^2(9+2\varepsilon)]^{1/3}}
\end{equation}
Now we are making some changes in the equation~(\ref{eq3}),
first of all replacing the deformation $\beta$ by $\varepsilon$.

One may assume \cite{bes61kd,bra72rmp,dam69np} that only the leading term of
the Hamiltonian, namely the potential written above, contributes essentially
to the derivative,
\begin{equation}
\frac{d H}{d \varepsilon} \simeq \frac{d V}{d \varepsilon}
\end{equation}
The contribution of  $P_{ij}$, denoted by $P_\varepsilon$ for a system with
one deformation coordinate, sometimes assumed  to be
neglijible small, will be discussed in the last section. 

The derivative is written as
\begin{equation}
\frac{1}{\hbar \omega _0^0}\frac{dV}{d\varepsilon } = \frac{3}{2} \left [
f_1(\varepsilon)\eta + f_2(\varepsilon)\xi^2 \right ]  
\end{equation}
in which
\begin{equation}
f_1=\frac{\varepsilon (\varepsilon
+6)+9}{[27-\varepsilon^2(9+2\varepsilon)]^{4/3}} 
\end{equation}
\begin{equation}
f_2=2\frac{\varepsilon (2\varepsilon
+3)-9}{[27-\varepsilon^2(9+2\varepsilon)]^{4/3}} 
\end{equation}
For a single deformation parameter the inertia tensor becomes a scalar
$B_\varepsilon$
whith a summation in eq.~\ref{eq3} performed for all states $\nu$, $\mu$
taken into consideration in the pairing interaction \cite{bol72pr}.

\subsection{Pairing interaction}

We consider a set of doubly degenerate energy levels $\{\epsilon_i\}$
expressed in units of $\hbar \omega_0^0$. Calculations for neutrons are
similar with those for protons, hence for the moment we shall consider only
protons. In the absence of a pairing field, the first $Z/2$ levels are
occupied, from a total number of $n_t$ levels available. Only few levels
below ($n$) and above ($n'$) the Fermi energy are contributing to the
pairing correlations. Usually $n'=n$. 
If $\tilde{g_s}$ is the density
of states at Fermi energy obtained from the shell correction
calculation $\tilde{g_s}=dZ/d\epsilon$, expressed in number of levels
per $\hbar \omega_0^0$ spacing, the level density is half of this
quantity: $\tilde{g_n}=\tilde{g_s}/2$.

We can choose as computing parameter, the cut-off energy (in units of
$\hbar \omega_0^0$), $\Omega \simeq 1 \gg \tilde{\Delta}$. Let us take
the integer part of the following expression
\begin{equation}
\Omega \tilde{g_s} /2 = n = n'
\end{equation}
When from calculation we get $n
> Z/2$ we shall take $n=Z/2$ and similarly if $n'>n_t -Z/2$ we
consider $n'=n_t -Z/2$.

The gap parameter $\Delta =|G|\sum_k u_kv_k$ and the Fermi energy
with pairing corellations $\lambda$ (both in units of $\hbar
\omega_0^0$) are obtained as solutions of a nonlinear system of two
BCS equations
\begin{equation}
n'-n = \sum_{k=k_i}^{k_f}\frac{\epsilon_k -\lambda}{\sqrt{(\epsilon
_k -\lambda)^2+\Delta^2}}
\end{equation}
\begin{equation}
\frac{2}{G} = \sum_{k=k_i}^{k_f}\frac{1}{\sqrt{(\epsilon
_k -\lambda)^2+\Delta^2}}
\end{equation}
where $k_i=Z/2-n+1$; $k_f=Z/2+n'$.

The pairing interaction $G$ is calculated from a continuous
distribution of levels
\begin{equation}
\frac{2}{G} = \int_{\tilde{\lambda} -\Omega}^{\tilde{\lambda} +\Omega}
\frac{\tilde{g}(\epsilon)d\epsilon}{\sqrt{(\epsilon -\tilde{\lambda})^2
+\tilde{\Delta^2}}}
\end{equation}
where $\tilde{\lambda}$ is the Fermi energy deduced from the shell
correction calculations and $\tilde{\Delta}$ is the gap parameter,
obtained from a fit to experimental data, usually taken as
$\tilde{\Delta}=12/\sqrt{A}\hbar\omega_0^0$. 
Both $\Delta_p$ and $\Delta_n$ decrease with increasing
asymmetry $(N-Z)/A$. 
From the above integral we get
\begin{equation}
\frac{2}{G} \simeq 2\tilde{g}(\tilde{\lambda})\ln \left (\frac
{2\Omega}{\tilde{\Delta}} \right )
\end{equation}
Real positive solutions of BCS equations are allowed if
\begin{equation}
\frac{G}{2}\sum_k \frac{1}{|\epsilon_k -\lambda|} > 1
\end{equation}
i.e. for a pairing force (G-parameter) large enough at a given
distribution of levels. The system can be solved numerically by
Newton-Raphson method refining an initial guess
\begin{eqnarray}
\lambda_0&=&(n_s\epsilon_d +n_d\epsilon_s)/(n_s+n_d)+ G(n_s-n_d)/2
\nonumber \\
\Delta_0^2&=&n_sn_dG^2 -(\epsilon_d -\epsilon_s)/4
\end{eqnarray}
where $\epsilon_s , n_s$ are the energy and degeneracy of the last
occupied level and $\epsilon_d , n_d$ are the same quantities for the
next level. Solutions around magic numbers, when  $\Delta \rightarrow
0$, have been derived by Kumar et al. \cite{kum77pr}.

As a consequence of the pairing correlation, the levels situated
below the Fermi energy are only partially filled, while those above
the Fermi energy are partially empty; there is a given probability
for each level to be occupied by a quasiparticle
\begin{equation}
v_k^2 = \frac{1}{2}\left [1-\frac{\epsilon_k -\lambda}{\sqrt{(\epsilon
_k -\lambda)^2+\Delta^2}} \right ]
\end{equation}
or a hole
\begin{equation}
u_k^2 = 1-v_k^2
\end{equation}
Only the levels in the near vicinity of the Fermi energy (in a range
of the order of $\Delta$ around it) are influenced by the pairing
correlations. For this reason, it is sufficient for the value of the
cut-off parameter to exceed a given limit $\Omega \gg
\tilde{\Delta}$, the value in itself having no significance.

\subsection{Variation with deformation}

The following relationship allows to calculate the effective mass in units
of $\hbar^2/(\hbar \omega_0^0)$
\begin{equation}
\frac{\hbar \omega_0^0}{\hbar^2} B_\varepsilon = \frac{9}{2}\sum_{\nu \mu} 
\frac{\langle \nu|f_1\eta+f_2\xi^2 |\mu \rangle \langle 
\mu|f_1\eta+f_2\xi^2|\nu
\rangle}{(E_\nu +E_\mu)^3}(u_\nu v_\mu +u_\mu v_\nu)^2
\label{eq35}
\end{equation}
Matrix elements are calculated by performing some integrals
\begin{eqnarray*}
\lefteqn{\langle n_z' n_r' m'|f_1(\varepsilon)\eta + f_2(\varepsilon)\xi^2| 
n_z n_r m \rangle = \delta _{m' m}N_{n_r'}^mN_{n_r}^mN_{n_z'}N_{n_z} 
\cdot [} \nonumber \\ & &
f_1 \int_0^\infty d\eta
\eta^{|m|+1}e^{-\eta}L_{n_r'}^{|m|}(\eta)L_{n_r}^{|m|}(\eta) 
\int_{-\infty}^{\infty} d\xi e^{-\xi^2}H_{n_z'}(\xi)H_{n_z}(\xi) + \\ & &
f_2 \int_0^\infty d\eta
\eta^{|m|}e^{-\eta}L_{n_r'}^{|m|}(\eta)L_{n_r}^{|m|}(\eta)
\int_{-\infty}^{\infty} d\xi \xi^2e^{-\xi^2}H_{n_z'}(\xi)H_{n_z}(\xi) ]
\end{eqnarray*}
where
\begin{equation}
N_{n_z'}N_{n_z}\int_{-\infty}^{\infty} d\xi e^{-\xi^2}H_{n_z'}(\xi)H_{n_z}(\xi) 
= \delta _{n_z' n_z}
\end{equation}
\begin{equation}
 N_{n_r'}^mN_{n_r}^m\int_0^\infty d\eta
\eta^{|m|}e^{-\eta}L_{n_r'}^{|m|}(\eta)L_{n_r}^{|m|}(\eta) = 
\delta _{n_r' n_r}
\end{equation}
so that
\begin{eqnarray*}
\lefteqn{\langle n_z' n_r' m'|f_1(\varepsilon)\eta + f_2(\varepsilon)\xi^2| 
n_z n_r m \rangle = \delta _{m' m}N_{n_r'}^mN_{n_r}^mN_{n_z'}N_{n_z} 
\cdot } \nonumber \\ & &
\left [\delta _{n_z' n_z} f_1 \int_0^\infty d\eta
\eta^{|m|+1}e^{-\eta}L_{n_r'}^{|m|}(\eta)L_{n_r}^{|m|}(\eta) 
+ \delta _{n_r' n_r}f_2 
\int_{-\infty}^{\infty} d\xi \xi^2e^{-\xi^2}H_{n_z'}(\xi)H_{n_z}(\xi) \right ]
\end{eqnarray*}
Next we can use the relationships \cite{web04}:
\begin{equation}
\int_0^\infty d\eta
\eta^{|m|+1}e^{-\eta}L_{n_r'}^{|m|}(\eta)L_{n_r}^{|m|}(\eta) =
\delta _{n_r' n_r}(2n_r + |m| + 1)\frac{(n_r+ |m|)!}{n_r!}
\end{equation}
\small{
\begin{equation}
\int_{-\infty}^{\infty} d\xi \xi^2e^{-\xi^2}H_{n_z'}(\xi)H_{n_z}(\xi) = 
\sqrt{\pi}n_z!2^{n_z} \left [\delta _{n_z' n_z}(n_z+\frac{1}{2}) + 
\delta _{n_z' n_z+2}(n_z+1)(n_z+2) + \delta _{n_z' n_z-2}\frac{1}{4} \right ]
\end{equation}}
which were obtained by using the recurrence relations and orthonormalization
conditions for Hermite polynomials and associated Laguerre polynomials.
\begin{equation}
L_n^k(x) = L_n^{k+1}(x) - L_{n-1}^{k+1}(x) \; \; ; \; \; 
(n+1)L_{n+1}^k(x)=[(2n+k+1)-x]L_n^k(x)-(n+k)L_{n-1}^k(x)
\end{equation}
\begin{equation}
H_{n+1}(x)=2xH_n(x) - 2nH_{n-1}(x)
\end{equation}
with particular values $L_0^k(x)=1$, $L_1^k(x)=1+k-x$, $H_0(x)=1$, $H_1(x)=2x$.

Eventually from the sum of equation (\ref{eq35}) one is left with an
important diagonal contribution and two nondiagonal terms
\begin{equation}
\frac{\hbar \omega_0^0}{\hbar^2} B_{\varepsilon 1} =
\frac{9}{4}\delta_{n_r' n_r}\delta_{m' m}\sum_{\nu =k_i}^{k_f} \left [
f_1(2n_r+|m|+1)+f_2(n_z+1/2) \right ]^2 \frac{(u_\nu
v_\nu)^2}{E_\nu^3}\delta_{n_z' n_z}
\label{eq42}
\end{equation}
\begin{equation}
\frac{\hbar \omega_0^0}{\hbar^2} B_{\varepsilon 2} =
\frac{9}{4}\delta_{n_r' n_r}\delta_{m' m}\sum_{\nu \ne \mu}
\frac{f_2^2}{2}(n_z+1)(n_z+2)\frac{(u_\nu
v_{\mu}+u_{\mu}v_\nu)^2}{(E_\nu+E_{\mu})^3}\delta_{n_z' n_z+2}
\label{eq43}
\end{equation}
\begin{equation}
\frac{\hbar \omega_0^0}{\hbar^2} B_{\varepsilon 3} =
\frac{9}{4}\delta_{n_r' n_r}\delta_{m' m}\sum_{\nu \ne \mu}
\frac{f_2^2}{2}(n_z-1)n_z\frac{(u_\nu
v_{\mu}+u_{\mu}v_\nu)^2}{(E_\nu+E_{\mu})^3}\delta_{n_z' n_z-2}
\label{eq44}
\end{equation}
where $k_i$ and $k_f$ have been defined above. In order to perform the
summations of the nondiagonal terms for a state with a certain $\nu$ 
(specified quantum numbers $n_z n_r m$) one has to consider only the states
with $\mu \ne \nu$ and $n_r'=n_r \ \ ; \ \ m'=m$ for which $n_z'=n_z+2$ or 
$n_z'=n_z-2$ respectively. Finally one arrives at the nuclear inertia in
units of  $\hbar ^2/$MeV by adding the three terms and dividing by 
$\hbar \omega_0^0$. 

\subsection{Hydrodynamical formulae}

There are several hydrodynamical formulae \cite{sob79ech} of the mass
parameters. For a spherical liquid drop with a radius $R_0=1.2249A^{1/3}$~fm
one has
\begin{equation}
B^{irr}(0)=\frac{2}{15}MAR_0^2=\frac{2}{15}
\frac{\hbar^2}{41.5\mbox{MeV$\cdot$fm$^2$}}1.2249^2A^{5/3}\mbox{fm$^2$}=
0.0048205A^{5/3} \; \frac{\hbar^2}{\mbox{MeV}} 
\label{eq45}
\end{equation}
When the spheroidal deformation is switched on it becomes 
\begin{equation}
B_\varepsilon^{irr}(\varepsilon) =B^{irr}(0)\frac{81}{[27-\varepsilon ^2(9
+2\varepsilon)]^{4/3}}\frac{9+2\varepsilon ^2}{(3-2\varepsilon)^2}
\label{eq46}
\end{equation}
Good results for the fission halflives of actinides have been obtained by
using an inertia larger by about an order of magnitude.

One can also employ a formula with a fit parameter $k$
\begin{equation}
B_r^{ph}(r) =\mu +k(B_r^{irr}(r) -\mu)
\end{equation}
where $k$ shows how much deviates the flow of nuclear matter from an
irrotational one. By substituting the above expression of the irrotational
term, one obtains
\begin{equation}
B_r^{ph}(r) =\mu \left [1+k\frac{17}{15}\exp \left (-\frac{r -3R_0
/4}{d} \right ) \right ]
\end{equation}
with parameter values of $d=d_0=R_0/2.452$ and $k=11.5$, or $d=2d_0$ and
$k=6.5$.

\section{Results}

The main result of the present study is represented by the equations
(\ref{eq42}--\ref{eq44}), which could be used to test complex computer codes
developed for realistic single-particle levels, for which it is not possible
to obtain analytical relationships.
\begin{figure}[htb]
\centerline{\includegraphics[height=10cm]{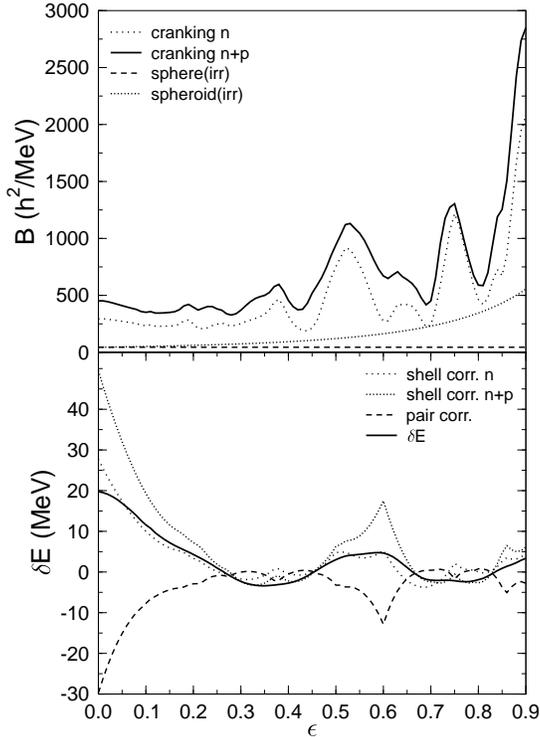}}
\caption{Top: comparison of the effective mass (in units of $\hbar
^2/$MeV) calculated by using the cranking model for the proton plus neutron
level schemes, only for neutrons, as well as for the irrotational spheroidal
and spherical shapes of $^{240}$Pu.
Bottom: shell corrections for neutrons and protons, only for neutrons,
pairing corrections, and shell plus pairing corrections.}
\label{fig1}
\end{figure}
Nuclear inertia of $^{240}$Pu calculated with the
equation~(\ref{eq45}) for a spherical liquid drop and with eq.~(\ref{eq46})
for spheroidal shapes is illustrated in figure~\ref{fig1}. 
One can see how $B^{irr}(0)$ increases when the mass number of the
nucleus is increased.
The irrotational value $B_\varepsilon^{irr}(\varepsilon)$ mo\-no\-to\-nously
increases with the spheroidal deformation parameter $\varepsilon$.
Due to the fact that in this single center model the nucleus only became
longer without developing a neck and never arriving to a scission
configuration when the deformation is increased, the reduced mass is not
reached as it should be in a two center model \cite{pg197pr95}.

The cranking inertia of the spheroidal harmonic oscillator
calculated by using the analytical
relationships (\ref{eq42}--\ref{eq44}) and the correction given in the next
section
shows very pronounced fluctuations which are correlated to the shell
corrections (calculated with the macroscopic-microscopic method
\cite{str67np}) plotted at the bottom of the figure~\ref{fig1}.

\subsection{Variation of the gap parameter and Fermi
energy with deformation}
\begin{figure}[htb]
\centerline{\includegraphics[width=7cm]{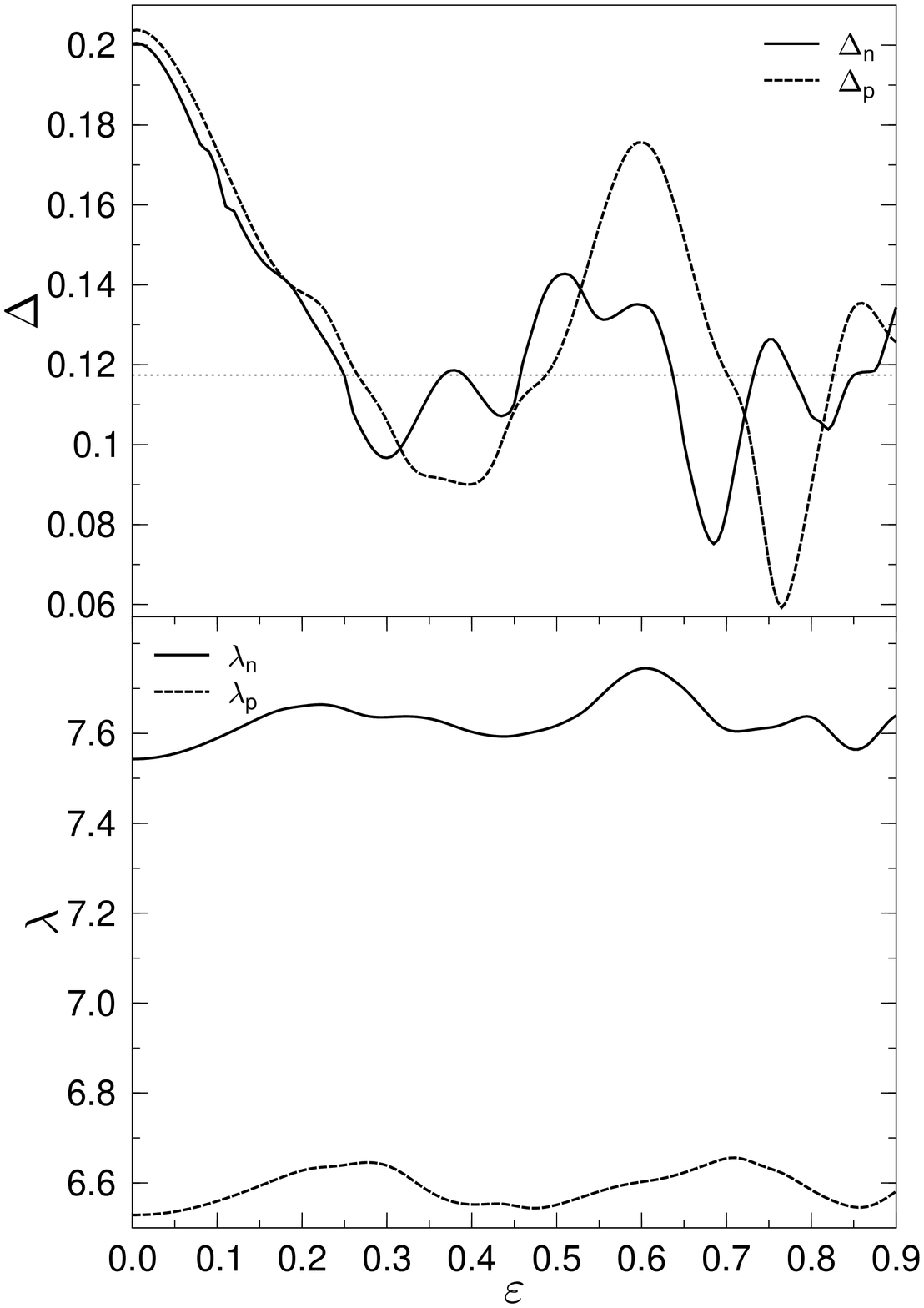}} 
\caption{ The variation with deformation 
of the solutions of BCS equations for Fermi energy $\lambda$ (bottom)
and the gap parameter $\Delta$ (top) 
of the proton and neutron level schemes for
$^{240}$Pu nucleus. The energies are expressed in units of $\hbar
\omega_0^0=6.597$~MeV. The dotted line in the upper part corresponds to
$\tilde{\Delta}=0.117$.}
\label{fig2}
\end{figure}

\begin{figure}
\centerline{\includegraphics[width=10cm]{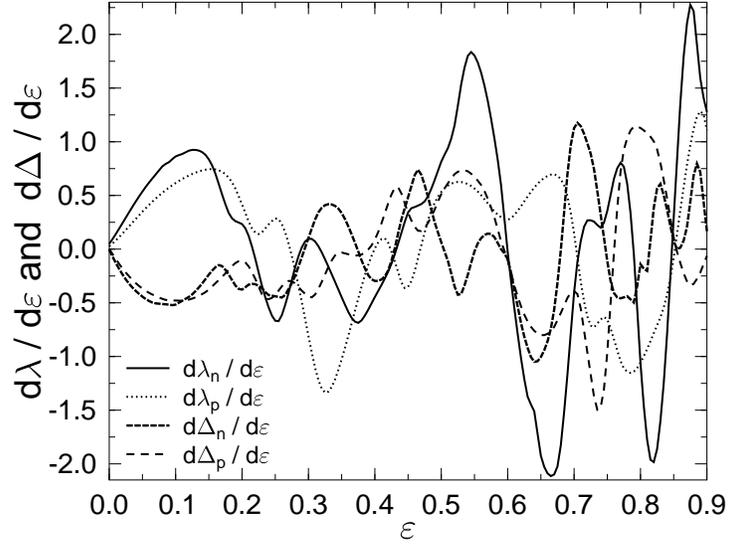}} 
\caption{
The derivatives with respect to deformation
of the solutions of BCS equations for Fermi energy $\lambda$
and the gap parameter $\Delta$ of the proton and neutron level schemes for
$^{240}$Pu nucleus. The energies are expressed in units of $\hbar
\omega_0^0=6.597$~MeV.}
\label{fig3}
\end{figure}
Now we can calculate the correction term as
\begin{eqnarray*}
\lefteqn{P_\varepsilon = \frac{2\hbar ^2}{8}\sum_\nu \frac{1}{E_\nu ^5} 
\left [ \left(\Delta\frac{d \lambda}{d \varepsilon} \right)^2
+ \right .} \nonumber \\ & &
(\epsilon_\nu - \lambda )^2\left(\frac{d \Delta}{d \varepsilon}
\right)^2 + 2\Delta(\epsilon_\nu - \lambda )\frac{d \lambda}{d
\varepsilon} \frac{d  \Delta}{d \varepsilon} \\ & &
-  2\Delta ^2 \frac{d \lambda}{d\varepsilon} \langle \nu | 
dV/d\varepsilon |\nu \rangle \\ & & \left .
-  2\Delta (\epsilon_\nu - \lambda )\frac{d \Delta}{d
\varepsilon} \langle \nu |dV/d\varepsilon | \nu \rangle \right ]
\end{eqnarray*}
\begin{figure}
\centerline{\includegraphics[width=10cm]{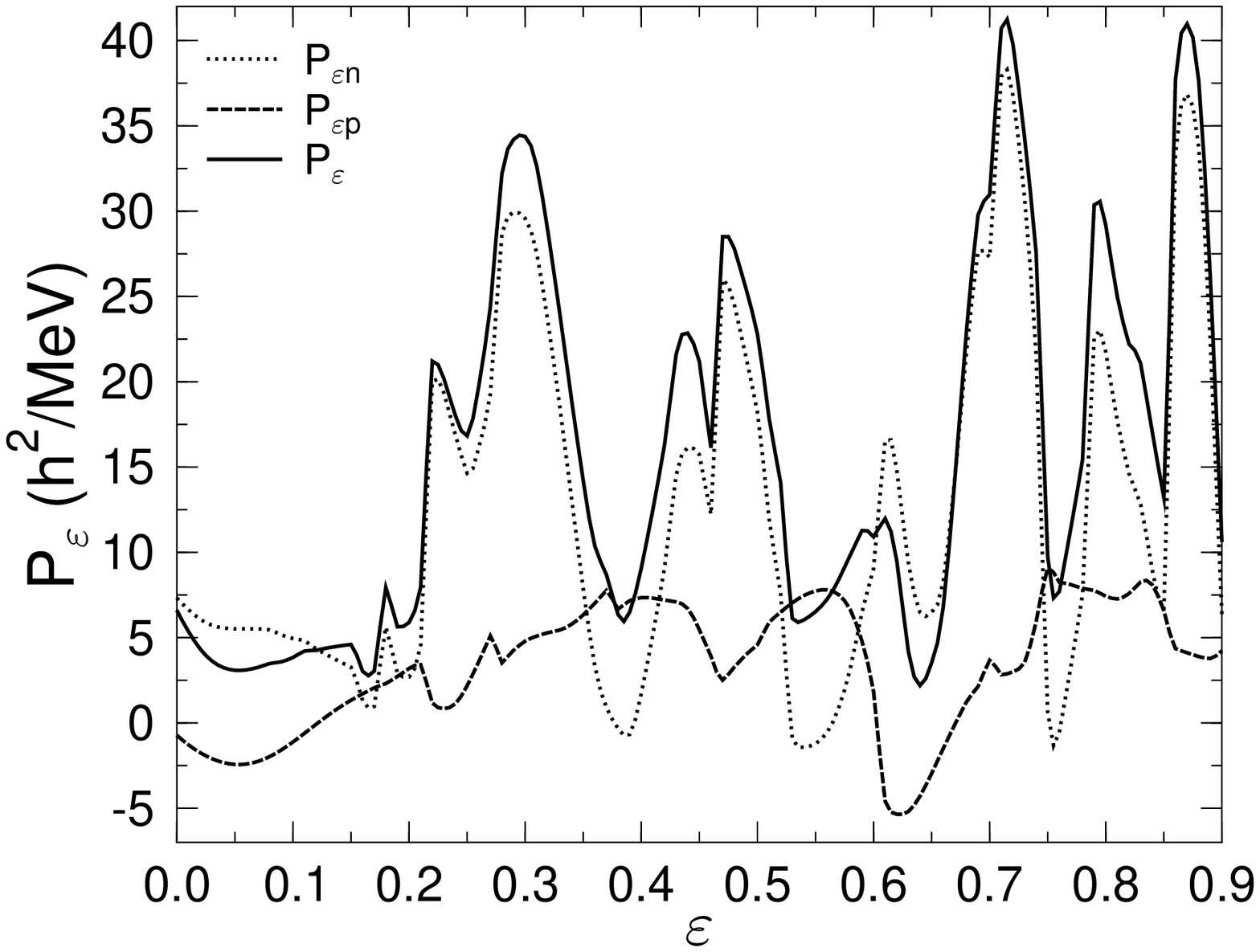}} 
\caption{Contribution, $P_\varepsilon$, to the mass parameter of the occupation 
number variation with deformation for $^{240}$Pu nucleus 
expressed in units of $\hbar^2/$MeV.}
\label{fig4}
\end{figure}
In figure \ref{fig2} we plotted the variation with deformation
of the solutions of BCS equations for Fermi energy $\lambda$ (bottom)
and the gap parameter $\Delta$ (top)
of the proton and neutron level schemes for
$^{240}$Pu nucleus. The dotted line at the value 0.117 corresponds to 
$\tilde{\Delta}$. Their derivatives with respect to the deformation
parameter are given in figure \ref{fig3}.
For superdeformed nuclei with $\varepsilon > 0.5$ the oscilllation
amplitudes of $d \lambda_n / d \varepsilon$ are approaching
their maximum values of about 2 units. In the same range of the deformations
the inertia is also larger as a result of the increased density of levels at
the Fermi surface.

The result displayed in figure \ref{fig4} shows the important contribution
of the neutron level scheme, $P_{\varepsilon n}$ (dotted line), reflecting
the larger density of states at the Fermi energy, compared to the proton
term $P_{\varepsilon p}$ (dashed line). Their sum is a positive quantity,
contributing to an increase of the nuclear inertia.
In a dynamical investigation using
the quasiclassical WKB approximation, the quantum tunnelling penetrability
depends exponentially on the action integral, in which the integral contains
a square root of the product of mass parameter and deformation energy. This
exponential dependence amplifies very much any variation of the inertia.
Consequently, the term $P_{ij}$ should be
considered in calculations. A similar conclusion was drawn from a study
of a realistic two-center shell model \cite{sch86zp}. 

\section{Conclusions}

By using the wave functions of the spheroidal harmonic oscillator (the
simplest variant of the Nilsson model) without spin-orbit coupling one can
obtain analytical relationships for the cranking inertia. Consequently the
result may be conveniently used to test complex computer codes  developed
for realistic two center shell models.

Unfortunately this single center oscillator is not able to describe fission
processes reaching the scission configuration or ground states with
necked in or diamond shapes. When the deformation parameter is increased the
nucleus became longer and longer without developing a neck and reaching the
touching point configuration. In this way it is not possible to obtain at
the limit a nuclear inertia equal to the reduced mass of the final fragments
in a process of fission, alpha decay or cluster radioactivity.

As expected, in agreement with the results obtained by other authors, the
cranking inertia is larger than the hydrodynamical one for a spheroidal
shape which is higher than that of a spherical nucleus.

{\bf Acknowledgment}
This work was partly supported 
by Deutsche Forschungsgemeinschaft, Bonn, and by
the Ministry of Education and Research, Bucharest.
We acknowledge also the support by Prof. S. Hofmann, Gesellschaft
f\"{u}r Schwerionenforschung (GSI), Darmstadt.

\end{document}